\documentclass[twoside,superscriptaddress,twocolumn,a4paper,pra,aps]{revtex4-1}

\usepackage{color}
\usepackage{graphicx}
\usepackage[utf8x]{inputenc}
\usepackage[T1]{fontenc}
\usepackage{siunitx}
\usepackage{hyperref}
\usepackage{tabularx}
\usepackage{filecontents}

\begin{document}

\title{Experimental construction of a versatile four-photon source}

\author{Karol Bartkiewicz} \email{bartkiewicz@jointlab.upol.cz}
\affiliation{Faculty of Physics, Adam Mickiewicz University,
PL-61-614 Pozna\'n, Poland}
\affiliation{RCPTM, Joint Laboratory of Optics of Palacký University and Institute of Physics of Czech Academy of Sciences, 17. listopadu 12, 771 46 Olomouc, Czech Republic}

\author{Antonín Černoch} \email{antonin.cernoch@upol.cz}
\affiliation{RCPTM, Joint Laboratory of Optics of Palacký University and Institute of Physics of Czech Academy of Sciences, 17. listopadu 12, 771 46 Olomouc, Czech Republic}
   
\author{Karel Lemr}
\thanks{Presently on leave at Faculty of Physics, Adam Mickiewicz University, Poland}
\email{k.lemr@upol.cz}
\affiliation{RCPTM, Joint Laboratory of Optics of Palacký University and Institute of Physics of Czech Academy of Sciences, 17. listopadu 12, 771 46 Olomouc, Czech Republic}

\begin{abstract}
The paper discusses technical aspects of constructing a highly versatile multi-photon source. The source is able to generate up to four photons which is sufficient for a large number of quantum communications protocols. It can be set to generate separable, partially-entangled or maximally-entangled photon pairs with tunable amount of purity. Furthermore, the two generated pairs can be prepared in different quantum states. In this paper, we provide all the necessary information needed for construction and alignment of the source. To prove the working principle, we also provide results of thorough testing.
\end{abstract}

\date{\today}

\maketitle
\section{Introduction}
Light is an outstanding carrier of information. For this reason, linear optics is one of the most heavily exploited platforms for quantum information processing (QIP) \cite{bib:nielsen:quantum_comput,bib:alber:quantum_inform}. Although many linear-optical quantum gates are probabilistic in their nature, they are experimentally well accessible \cite{bib:kok:linear}. These particular qualities make linear optics an appealing platform for proof-of-principle experiments \cite{bib:bouwmeester:teleport,bib:brien:cnot,bib:kiesel:cphase,scarani2005cloning,soubusta2007severalcloners} and for implementing quantum communications protocols \cite{bib:bb84:exper,ralph99CVcrypto,marcikic200450km,Ma2012,bib:lemr:router2}. Both continuous and discrete states of light have been used to transmit quantum information. The latter make use of individual photons and encode qubits into such degrees of freedom as polarization, position or orbital angular momentum \cite{Bliokh2015,BLIOKH20151,Lodahl2017}.

There is a broad portfolio of physical processes that lead to single-photon emission or at least to emission of states close to the Fock $|1\rangle$ state. The single-photon sources reported in the litterature are based on single atoms \cite{Hijlkema2007,PhysRevLett.89.067901}, ions \cite{1367-2630-6-1-094,Keller2004,1367-2630-11-10-103004}, molecules \cite{PhysRevLett.83.2722,PhysRevA.58.620}, quantum dots \cite{Senellart2017,Aharonovich2016,Kako2006,Santori2002} or even crystal defects \cite{PhysRevLett.85.290,Brouri:00,PhysRevA.64.061802}. While these are promising techniques, in the majority of linear-optical QIP experiments so far, photons were generate in the process of spontaneous parametric down-conversion (SPDC) in bulk crystals \cite{PhysRev.124.1646,bib:zeldocich:69,PhysRevLett.25.84,PhysRevLett.100.133601,PhysRevA.66.053805} or waveguides \cite{PhysRevLett.93.093601,doi:10.1063/1.3132086,Karpinski:12,PhysRevA.87.013836,Jachura:14}.

SPDC is a non-linear optical process occurring in materials with second order optical non-linearity \cite{boyd2003nonlinear}. This process results in spontaneous transformation of a pumping photon into a pair of photons reffered to as signal and idler. Mathematically, it can be described in terms of an interaction Hamiltonian (in the rotating frame) \cite{PhysRevA.50.5122}
\begin{equation}
\hat{H}_\mathrm{int} = \chi \hat{a}_s^\dagger \hat{a}_i^\dagger \hat{a}_p + \mathrm{h.c.},
\end{equation}
where $\chi$ in an interaction constant, $\hat{a}_s^\dagger$ and $\hat{a}_i^\dagger$ stand for creation operators of the signal and idler modes and $\hat{a}_p$ is the annihilation operator of the pumping mode. In typical configuration, the pumping mode is in a non-depletable coherent state (strong laser pumping) and the interaction is sufficiently weak to approximate the states of signal and idler modes using an expansion in the Fock basis
\begin{equation}
\label{eq:state}
|\psi_{si}\rangle \propto |00\rangle + \kappa |11\rangle + \frac{\kappa^2}{2} |22\rangle + ...,
\end{equation}
where $\kappa = \frac{it}{\hbar}\chi\langle\hat{a}_p\rangle$ and $t$ is the interaction time. Given that $\kappa \ll 1$, vacuum becomes the predominant term in Eq. (\ref{eq:state}). To increase the overlap of the state in Eq.~(\ref{eq:state}) with a perfect Fock $|1\rangle$ state, SPDC-based single-photon sources are often used in conjunction with detection post-selection \cite{PhysRevA.66.024308} or heralding \cite{PhysRevLett.56.58}. For instance, results are registered only if photons are detected in both signal and idler modes simultaneously. This strategy allows to effectively reduce the generated state into
\begin{equation}
\label{eq:state_post-select}
|\psi_{si}\rangle \propto \kappa |11\rangle + \frac{\kappa^2}{2} |22\rangle + ...,
\end{equation}
which can be made arbitrarily close to a perfect Fock state $|11\rangle$ by decreasing the interaction strength or pumping beam power. The benefit of SPDC-based light sources is that two photons in different spatial modes are generated at the same time and can, thus, be immediately used in two-qubit QIP logical gates.
\begin{center}
\begin{figure*}
\includegraphics[scale=1]{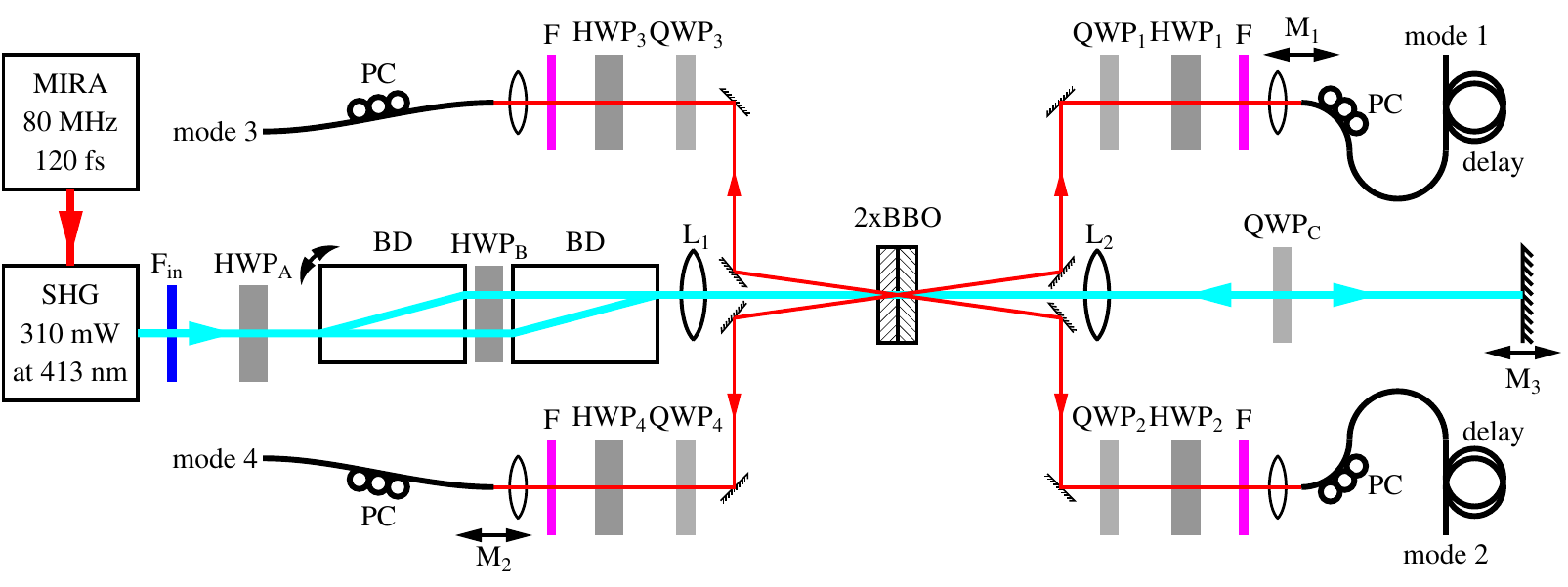}
\caption{\label{fig:setup} Schematic representation of the four-photon source. Individual components are labeled as follows: BD -- beam displacer, HWP -- half-wave plate, QWP -- quarter-wave plate, L -- lens, BBO -- $\beta$-barium borate, F -- filter, M -- motorized translation, PC -- polarization controller. Working principle is described in the text.}
\end{figure*}
\end{center}

Historically, SPDC has been used as a two-photon source in the seminal Hong-Ou-Mandel  experiment \cite{bib:hom} or to power some of the key experiments including fundamental quantum gates \cite{bib:brien:cnot,bib:kiesel:cphase,scarani2005cloning,bib:lanyon:simplify} or quantum communications protocols \cite{Ma2012,bib:kocsis:amplifier,soubusta2008cloning}. SPDC not only allows to generate two individual photons at once, but it also allows to generate these photons in an entangled state. Originally, entanglement was achieved by using Type II SPDC process \cite{PhysRevLett.75.4337}, like for instance in the first teleportation experiment \cite{bib:bouwmeester:teleport}. Recently, however, a crystal cascade (a.k.a. Kwiat source) becomes more and more popular \cite{bib:kwiat:crossed} because of its versatility. In this configurations, two crystals cut for Type~I SPDC process are placed so that their axes are in mutually perpendicular planes. Thus, the crystals generate a photon pair with mutually orthogonal polarizations. Entanglement arises from the coherent generation of photons in these two crystals. 

Experiments requiring more than two photons rely on repeated generation of photon pairs \cite{PhysRevA.90.042302,Patele1501531,PhysRevLett.117.210502}, remarkably even with independent sources of pumping \cite{PhysRevA.79.040302}. It was shown that the process can scale up to 5 generated photon pairs \cite{PhysRevLett.117.210502}. To minimize generation-time jitter across the generated pairs, ultrashort pumping pulses are applied. The combination of crystal cascade and femtosecond pulsing was experimentally demonstrated in 2002 but with only one generated photon pair \cite{PhysRevA.66.033816}. Later, Dobek \emph{et al.} generalized the scheme to obtain two pairs (four photons) \cite{PhysRevLett.106.030501,1555-6611-23-2-025204}. Their setup, however, does not allow to generate the two photon pairs in mutually different states.

In this paper, we describe in detail construction of a highly versatile four-photon source. To our best knowledge, this configuration brings more tunability and, in comparison to previously reported sources, allows to generate a broader class of quantum states. By using the pumping beam twice, our source is more effecient than other recent designs \cite{PhysRevA.93.062329}.

\section{Technical description}
\label{sec:tech}
The core of our four-photon source is a BBO ($\beta$-barium borate) crystal cascade cut for Type~I SPDC (see Fig.~\ref{fig:setup}). The crystal cascade consists of two crystals, each \SI{0.1}{\milli\meter} thick, optically contacted so that their optical axes lie in perpendicular planes (manufactured by Cleveland crystals, USA). Cut angles for these crystals are 29.1 degrees resulting in an angle of about 4 degrees between propagation directions of the pump beam and the generated photons. One of the crystals converts horizontally polarized pump beam into vertically polarized photon pairs ($|VV\rangle$). In the second crystal, horizontally polarized photon pairs ($|HH\rangle$) are obtained from vertically polarized pumping photons. Note that these two processes are indistinguishable and coherent. Thus, when the cascade gets pumped by a generally polarized light beam, polarization entangled state
\begin{equation}
\label{eq:ent_state}
|\psi_\mathrm{si}\rangle = \cos\alpha |HH\rangle + \mathrm{e}^{i\varphi}\sin\alpha|VV\rangle
\end{equation}
is produced, where $\tan{\alpha}$ corresponds to the ratio between amplitudes in the horizontal and vertical polarization modes of the pumping beam and $\varphi$ to their mutual phase shift. For more details about the working principle of the crystal cascade, see the original proposal in Ref.~\cite{bib:kwiat:crossed}.

The pumping beam originates from a Coherent Mira laser system which generates 80 million femtosecond pulses per second spectrally centered at \SI{826}{\nano\meter} with spectral full width at half of maximum (FWHM) of about \SI{10}{\nano\meter} and duration of about \SI{120}{\femto\second}. Typical output mean power measured directly behind the laser was \SI{850}{\milli\watt}. Next, the beam gets frequency doubled in a second-harmonics generation (SHG) unit (home-made) equipped with a \SI{2}{\milli\meter} thick BBO crystal. Now, the frequency-doubled beam centered at \SI{413}{\nano\meter} has spectral width of \SI{4.4}{\nano\meter} and typically \SI{310}{\milli\watt} of optical power. At the output of the SHG unit the remaining light of the fundamental beam was filtered out by narrow-band interference filter F$_{\rm in}$.

In the next step, we tailor the pumping beam properties. A half-wave plate (HWP$_\mathrm{A}$) is used to change the ratio between horizontal and vertical polarization component of the beam and thus tune the angle~$\alpha$ of the generated state (\ref{eq:ent_state}). As a result of group velocity polarization dispersion that occurs in the BBO material \cite{PhysRevA.66.033816}, the horizontally and vertically polarized wave packets of the generated photons are shifted in time. This effect is comparable to the time duration of the femtosecond pumping pulses and has to be compensated for. We implement such compensation in a form of a polarization Mach-Zehnder interferometer consisting of two beam displacers (BD40 by Thorlabs) with a half-wave plate placed between them. The HWP is rotated by 45 degrees with respect to horizontal polarization direction. By unbalancing the interferometer arms, we mutually delay the horizontal and vertical polarization components of the pumping beam. Such shift can be set so that it compensates the subsequent polarization dispersion in the BBO material. The $|HH\rangle$ and $|VV\rangle$ components of the generated photon pair then become indistinguishable in time. On the other hand, the compensation can be made deliberately incorrect resulting in generation of pairs with variable purity. Using a piezo element for tilting one of the beam displacers, we are also able to change the mutual phase shift between the interferometer arms and thus tune the parameter $\varphi$ of the generated state (\ref{eq:ent_state}). Two lenses L$_1$ and L$_2$ with focal length \SI{150}{mm} and \SI{180}{mm} respectively were used to focus beam into the crystal cascade in forward and backward direction.

The properly tailored pumping beam impinges the crystal cascade and photons are generated into the forward propagating modes 1 and 2. Then the pumping beam gets reflected on the mirror mounted on motorized translation M$_3$ and re-enters the crystal cascade in the backward direction generating photons into modes 3 and 4. We place a quarter-wave plate QWP$_\mathrm{C}$ between the crystal cascade and the mirror so that polarization of the pumping beam can be fully or partially swapped (horizontal $\leftrightarrow$ vertical). This way we can fully or partially compensate for the polarization dispersion. As a result, we generate a quantum state in the backward propagating direction with different purity from the state in the forward direction. Note that because the pumping beam passes this QWP twice, it acts effectively as a HWP.

The photons travel about \SI{20}{\centi\meter} from the crystals in free space before being collected by fiber couplers equipped with narrow interference filters F (various spectral widths -- see testing below) into single-mode fibers. Hence, we implement both spectral and transversal mode filtering. By finetuning the position and focus on these couplers, we can collect photons predominantly from one of the crystals. Thus we control the parameter~$\alpha$ of the collected state. We were for instance able to generate a maximally entangled state in the forward direction and then only couple signal from one crystal in the backward direction obtaining there a separable state. We can also place HWPs and QWPs in front of the fiber couplers to transform the generated photonic states even further (e.g., to generate a singlet Bell state) or vary the angle $\varphi$ by tilting these plates. Fibers are equipped with polarization controllers which negate polarization changes in non-polarization maintaining fibers. Modes in forward direction are delayd by \SI{1}{m} of additional fiber to be coincident in time with backward modes.

\section{Alignment}
\label{sec:align}
In this section, we review the key steps involved in the alignment of the four-photon source. One starts by adjusting the laser system and subsequent SHG unit. These procedures however depend on the specific device used. Thus, we refer the reader to the relevant user's manual.

To achieve a reproducible beam propagation through the setup each time the laser is turned on, we insert several irises to the pumping beam path. As a first step, all lenses are removed and by steering mirrors at the output of the SHG unit, the beam is traced through the irises all its way to the final mirror M$_3$ (we recommend placing temporarily a beam block in front of M$_3$). Lenses are then placed back and by means of their transversal positioning, one makes sure that the forward propagating beam is also well aligned with the irises. Finally, tilts of the mirror M$_3$ are used to trace the reflected beam through the irises and back close to the output of the SHG unit.

All fiber couplers are connected to single-photon detectors (avalanche photodiodes) and their tilts and positions are optimized to couple as much signal as possible. We usually start the procedure with edge filters and once a strong signal is observed, we switch to narrower interference filters. Using proper electronics (e.g., TAC-SCA modules by Ortec with \SI{5}{\nano\second} windows), we register the two-fold coincident detections between photon couplers corresponding to the forward and backward propagating photon pairs. To maximize these coincidences, we gradually change the vertical tilt and position of one of the fiber coupler of each photon pair.

To generate entangled photon pairs of a given purity, the wave plates HWP$_\mathrm{A}$ and QWP$_\mathrm{C}$ are set to modify the pumping beam polarization as described in the previous section. Purity of the forward generated entanglement is tuned by vertically tilting one of the beam displacers (BD) thus introducing polarization dispersion to compensate for the polarization dispersion inherent to BBO.

Finally, we detect the four-fold coincidences in the following way: first, we use TAC-SCA modules (\SI{5}{\nano\second} windows) to register coincident detections between photons in modes 1\&3 and 2\&4 (photons of different pairs). Subsequently, we obtain the four-fold coincident detections by joining the outputs from TAC-SCA units on a coincidence logic with a wide coincidence window of \SI{460}{ns}. Due to the low rate of coincident generation of the two pairs, this broad window does not introduce significant amount of accidental coincidences.

In our particular case, a properly aligned setup was able to generate single photons with generation yields ranging typically from 10 to \SI{100}{\kilo\hertz} depending on the spectral filter used. Forward and backward propagating photon pairs occur with frequency 1 to \SI{10}{\kilo\hertz}. Frequency of coincident detection between two photons of different pairs is 10 to \SI{100}{\hertz}. Four photon coincident detections occur 1 to 10 times per minute.

\section{Testing}
\label{sec.test}
\subsection{Output state tomography}
\begin{figure}
\includegraphics[width=0.235\textwidth]{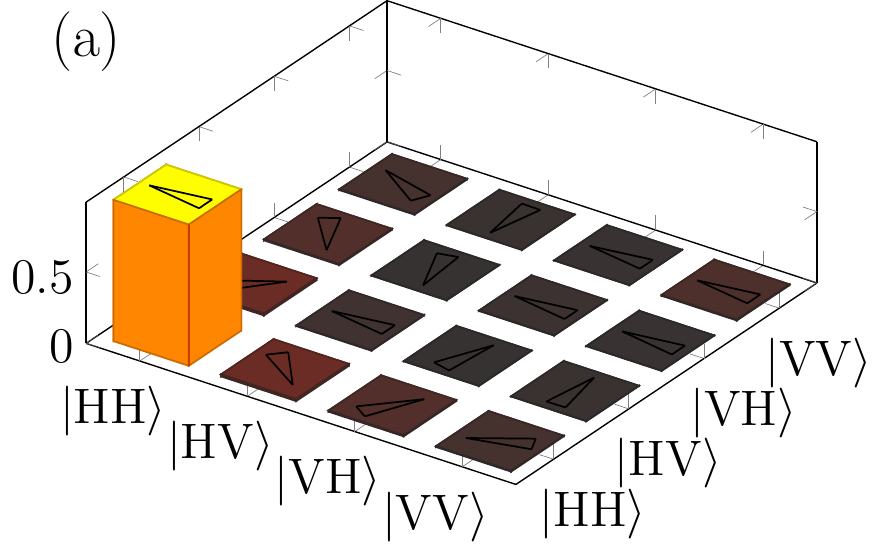}
\includegraphics[width=0.235\textwidth]{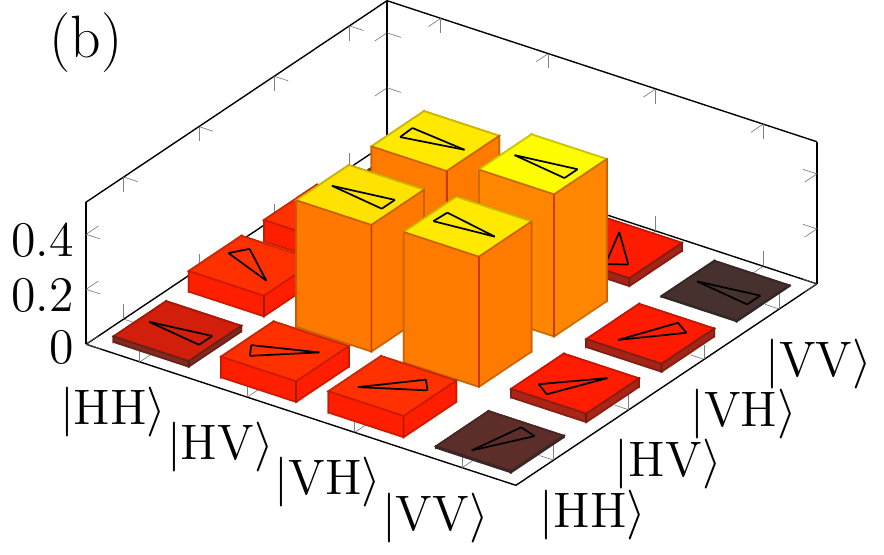}\\
\includegraphics[width=0.235\textwidth]{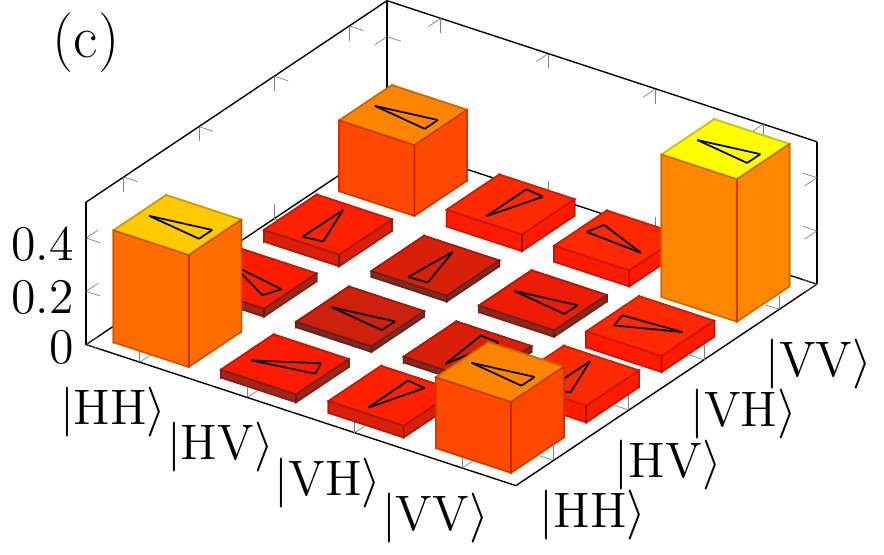}
\includegraphics[width=0.235\textwidth]{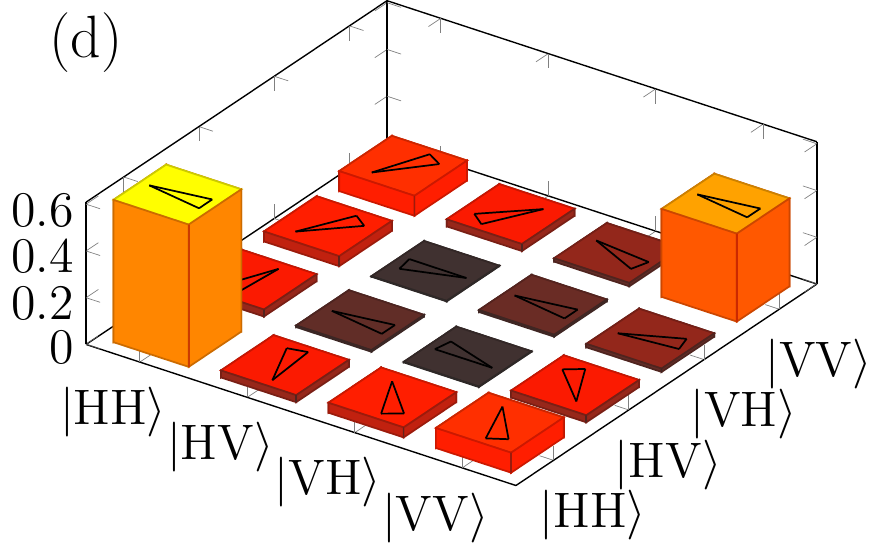}
\caption{\label{fig:qst} Density matrices of four representative two-photon quantum states: (a) separable state $|HH\rangle$, (b) singlet Bell state $\frac{1}{\sqrt{2}}\left(|HV\rangle-|VH\rangle\right)$, (c) $\hat{\rho}_\mathrm{si}(p=0.75)$ and (d) $\hat{\rho}_\mathrm{si}(p=0.50)$. States (c) and (d) are defined in Eq. (\ref{eq:mixed}). Bars denote amplitudes of the density matrix elements, arrows indicate the phase of each element.}
\end{figure}

We have subjected our source to a series of tests to verify that it operates correctly. In the first step, we have implemented a full quantum state tomography of the forward and backward propagating photon pairs. The tomography consists of performing combinations of polarization projection measurements accomplished by a series of HWP, QWP and a polarizer in front of each fiber coupler. For more details about this procedure, refer to Ref.~\cite{halenkova12detector}. Simultaneously, we have registered coincident detections of photons in modes 1 and 2 (for the forward propagating pair tomography) and modes 3 and 4 (backward propagating pair tomography). For this particular measurement, the fiber couplers were equipped with \SI{5.5}{\nano\meter} interference filters. We have carried out quantum state tomography on a wide range of source settings, observing pure, mixed, separable and entangled states. Table \ref{tab:qst} summarizes parameters of the four most representative states. The respective density matrices are depicted in Fig.~\ref{fig:qst}. Mixed states presented are in the form of
\begin{equation}
\label{eq:mixed}
\hat{\rho}_\mathrm{si}(p) = p|\Phi^+\rangle\langle\Phi^+| + (1-p)|\Phi^-\rangle\langle\Phi^-|,
\end{equation}
where $|\Phi^\pm\rangle = \frac{1}{\sqrt{2}}\left(|HH\rangle\pm|VV\rangle\right)$. We have obtained comparable results for both forward and backward propagating photon pairs. Typically, we are able to reach purities above 95\% even for maximally entangled states which are the most sensitive to decoherence. The source also allows to generate maximally mixed states of purities close to 50\%.
\begin{table}
\caption{\label{tab:qst}Parameters of four representative states generated by the source. Fidelity \cite{Jozsa94} is calculated with respect to the targeted quantum state. Mixed state are defined in Eq. (\ref{eq:mixed}).}
\begin{ruledtabular}
\begin{tabular}{llll}
Targeted state & Purity & Negativity & Fidelity\\\colrule
$|HH\rangle$       & 0.978$\pm$0.002 & 0.002$\pm$0.005 & 0.994$\pm$0.001\\
$\frac{1}{\sqrt{2}}\left(|HV\rangle-|VH\rangle\right)$   & 0.954$\pm$0.014 & 0.473$\pm$0.007 & 0.981$\pm$0.004\\
$\hat{\rho}_\mathrm{si}(p=0.75)$ & 0.627$\pm$0.016 & 0.253$\pm$0.013 & 0.975$\pm$0.004 \\
$\hat{\rho}_\mathrm{si}(p=0.50)$ & 0.534$\pm$0.005 & 0.083$\pm$0.009 & 0.981$\pm$0.002
\end{tabular}
\end{ruledtabular}
\end{table}

\begin{figure}
\includegraphics[scale=1]{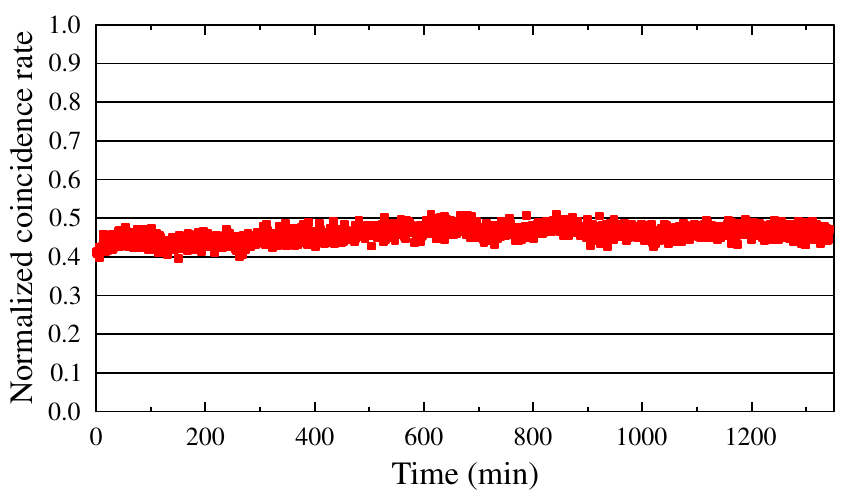}
\caption{\label{fig:stability} Stability test measuring the relative rate of detected coincidences across several hours. For details, see the text.}
\end{figure}

Due to the relatively low generation yield, multi-photon quantum experiments usually take long time to acquire enough signal. It is crucial that the source remains stable during the entire measurement. We have tested the phase $\varphi$ stability by preparing a triplet Bell state $|\Phi^+\rangle$ in the forward direction and then performing polarization projection onto diagonal linear polarization $\frac{1}{\sqrt{2}}\left(|H\rangle+|V\rangle\right)$ in mode 1 and onto circular polarization $\frac{1}{\sqrt{2}}\left(|H\rangle+i|V\rangle\right)$ in mode 2. It is easy to see that we should obtain half of the coincidence rate with respect to signal observed when both modes are projected onto diagonal polarization. If the phase $\varphi$ changes, the observed coincidence detection rate either increases or decreases. Thus, stability of the coincidence rate certifies stability of the phase $\varphi$. Figure \ref{fig:stability} depicts the evolution of observed coincidence rate in the above described configuration across several hours. The phase $\varphi$ remains stable within typical range of $\pm\pi/30$ during the entire measurement.

\subsection{Two-photon interference between photons of identical pair}
\begin{figure}
\includegraphics[scale=1]{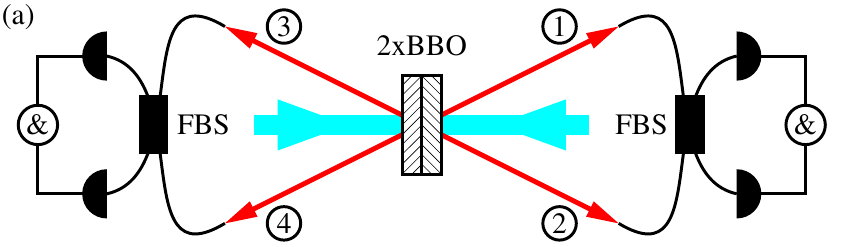}\\\vspace{1em}
\includegraphics[scale=1]{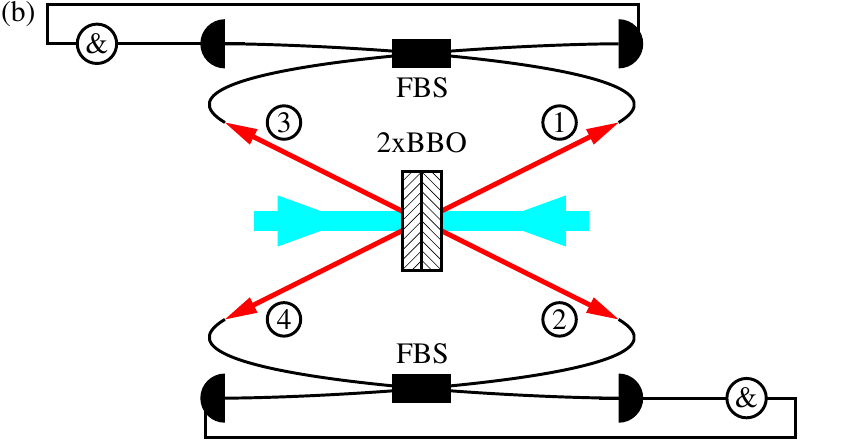}\\\vspace{1em}
\includegraphics[scale=1]{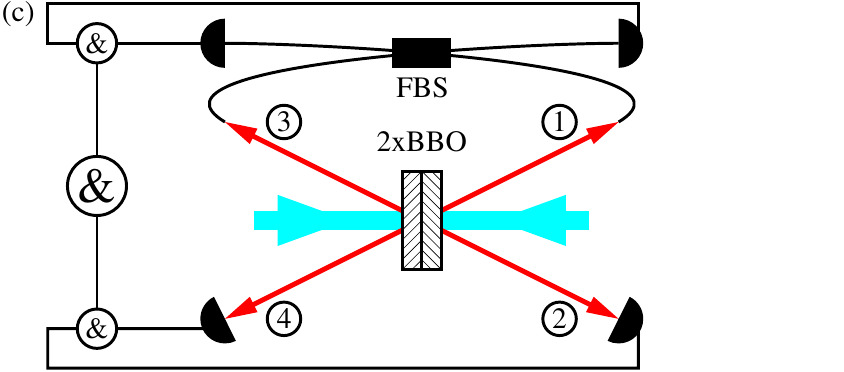}
\caption{\label{fig:setupHOMs} Configurations used in testing of two-photon interference between (a) photons belonging to a SPDC-generated pair, (b) photons of different pairs, and (c) heralded photons of different pairs. Components are labeled as follows: 2$\times$BBO -- crystal cascade, FBS -- fiber beam splitter, black semicircles -- detectors, \& -- coincidence registration. Photon modes are labeled by encircled numbers.}
\end{figure}
Multiphoton linear-optical quantum gates are basically complex single and two-photon interferometers. Carefully tailored interference then produces the desired information processing effect. Two-photon interference, or two-photon bunching, is a purely quantum effect. In order to observe high visibility interference patterns, the interfering photons must be (a) close to ideal Fock $|1\rangle$ states and (b) must be indistinguishable in all degrees of freedom (i.e., polarization, transversal mode, frequency spectrum). Moreover, the photons must also be indistinguishable in the arrival time to the detectors. This condition is quite elegantly fulfilled when dealing with photons generated in pairs as the jitter between their respective generation times is negligible.

To test the indistinguishability of the photons generated by our source, we have appended a simple two-photon interferometer to the forward and then also backward propagating pairs. This interferometer consists of a balanced 2$\times$2 fiber coupler with output ports leading to the detectors (see Fig.~\ref{fig:setupHOMs}a). It is well known that indistinguishable photons bunch on such fiber coupler resulting in lack of coincident detections observed by the detectors. To observe this bunching effect, we have equipped one of the fiber couplers with a motorized translation allowing to change the arrival time of one photon with respect to the other. The quality of interference and the indistinguishability is expressed in terms of visibility $V$ defined as
\begin{equation}
\label{eq:visibility}
V = \frac{cc_\mathrm{max}-cc_\mathrm{min}}{cc_\mathrm{max}+cc_\mathrm{min}},
\end{equation}
where $cc_\mathrm{max}$ and $cc_\mathrm{min}$ stand for coincidence detection rate in the maximum (photons arrive at different times) and minimum (photons arrive simultaneously) of the interference pattern. The exact shape of two-photon interference pattern as a function of arrival time difference strongly depends on spectral filtering applied by filters F. Typically, for Gaussian-shaped identically filtered photons, the interference pattern takes the form of the so-called Hong-Ou-Mandel dip which is a Gaussian shaped valley centered at zero arrival time difference.

\begin{figure}
\includegraphics[scale=0.5]{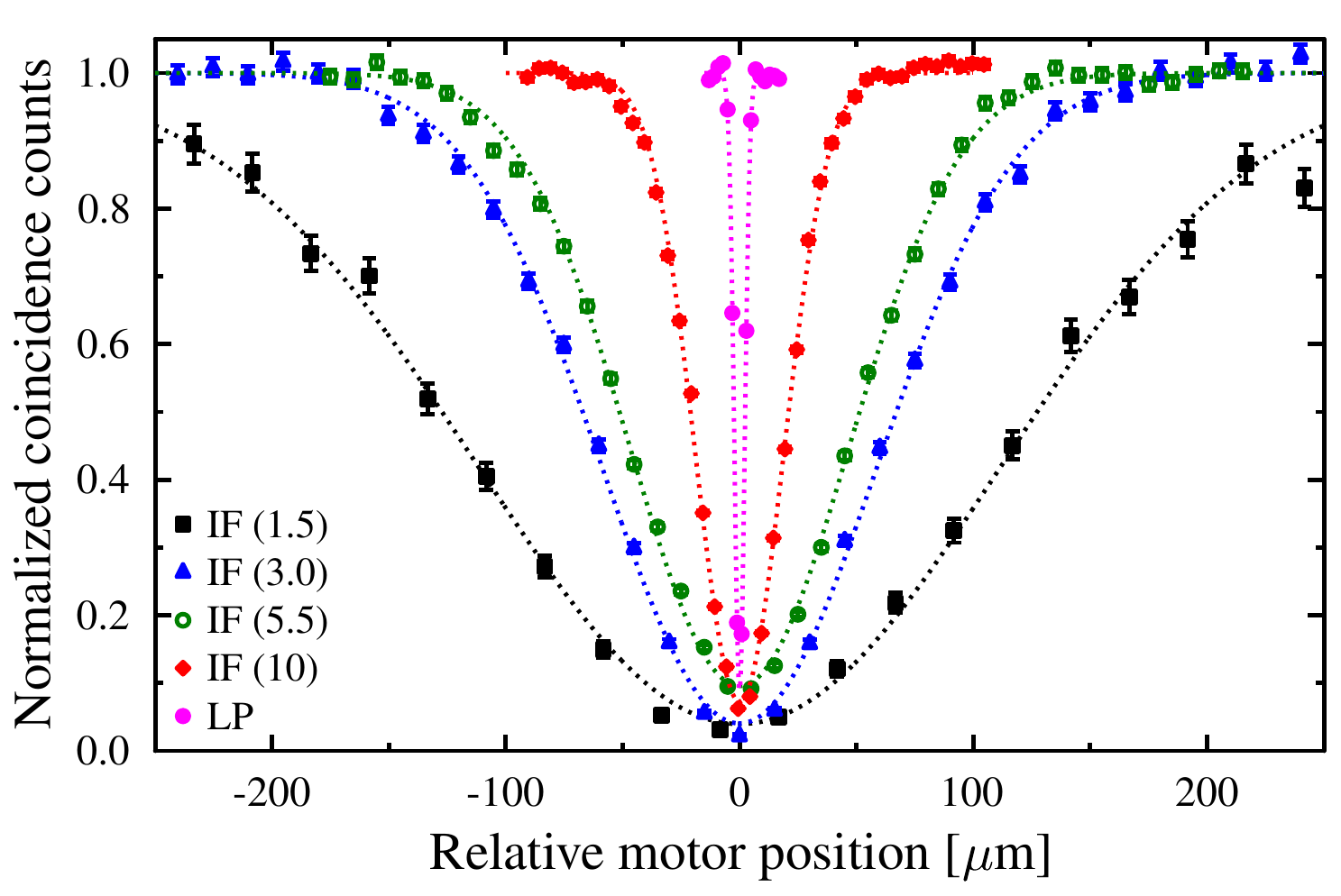}
\caption{\label{fig:HOMPF} Hong-Ou-Mandel dips observed on photons of the forward propagating pair using various filters as described in the caption of Table \ref{tab:homPF}.}
\end{figure}

First, we have tested the two-photon interference with several spectral filters on the forward propagating photon pair. The obtained Hong-Ou-Mandel dips are depicted in Fig.~\ref{fig:HOMPF} and their visibilities and widths are summarized in Table \ref{tab:homPF}. The presented values are raw data without any corrections on photo-pulse statistics. Dip widths are presented as FWHMs in spatial domain FWHM$_s$ (distance traveled by motorized translation stage), time domain FWHM$_t$ (corresponding temporal delay), and in spectral domain FWHM$_f$ using the relation
\begin{equation}
{\rm FWHM}_f = \frac{2 \sqrt{2} \ln{2}}{\pi} \frac{\lambda^2}{\rm FWHM_s}
\end{equation}
with $\lambda = \SI{826}{\nano\meter}$ being the central wavelength. Note that ideally, FWHM$_f$ should coincide with the FWHM of the filter used.
\begin{table}
\caption{\label{tab:homPF} Visibility $V$ and dip widths observed on on photons of the forward propagating pair. Filters are denoted: LP -- long-pass edge filter (\SI{780}{\nano\meter} edge), IF(FWHM) -- interference filters of given FWHM in units of \si{\nano\meter}.}
\begin{ruledtabular}
\begin{tabular}{lrrrr}
Filter used& FWHM$_s$& FWHM$_t$& FWHM$_f$& $V$ \\
 & (\si{\micro\meter}) & (\si{\femto\second}) & (\si{\nano\meter}) &  \\ \colrule\\
LP  		& 5.2$\pm$0.1 	& 17.3$\pm$0.3 	& 82.0$\pm$0.2 	& 0.663$\pm$0.001 \\
IF (10)& 44.5$\pm$0.2 	& 148.3$\pm$0.7	& 9.6$\pm$0.1 	& 0.830$\pm$0.003 \\
IF (5.5)& 111.1$\pm$0.6	& 370.3$\pm$2.0	& 3.8$\pm$0.1 	& 0.807$\pm$0.005 \\
IF (3.0)& 138.4$\pm$1.0	& 461.3$\pm$3.3	& 3.1$\pm$0.1 	& 0.932$\pm$0.005 \\
IF (1.5)& 263.2$\pm$6.8	& 877$\pm$23		& 1.6$\pm$0.1 	& 0.911$\pm$0.012 \\ 
\end{tabular}
\end{ruledtabular}
\end{table}

In the next step, we have repeated the same interference measurement on the backward propagating pair. The obtained results were comparable to those presented for the forward propagating pair. For instance visibility observed with IF (3.0) in forward direction was $V=0.932\pm0.005$ while in backward direction it reads $V = 0.910\pm0.005$. We conclude this testing step by noting that the source allows to generate pairs of mutually indistinguishable photons. This is manifested by reaching visibilities above $90\%$ when sufficient filtering is used.

\subsection{Two-photon interference between entangled photons}
Although the combination of quantum state tomography and two-photon interference presented in the previous subsection fully characterizes the emitted photon pairs, we also apply another diagnostic method, i.e., the two-photon interference measurement on entangled photon pairs. This method is of significant practical importance because it allows to quickly check for both entanglement quality and photon indistinguishability in one simple Hong-Ou-Mandel dip measurement. This method relies on the fact that interfering photons being initially in a singlet Bell state $|\Psi^-\rangle = \frac{1}{\sqrt{2}}\left(|HV\rangle - |VH\rangle\right)$ anti-bunch instead of bunching as seen for symmetric two-photon states. As a result, one does not observe a dip but an anti-dip which is a Gaussian-shaped peak in the coincident detection rate centered at position associated with zero temporal delay between the photons. Visibility of an anti-dip is defined using a modified visibility formula
\begin{equation}
\bar{V} = \frac{cc_\mathrm{max} - cc_\mathrm{min}}{3 cc_\mathrm{min}-cc_\mathrm{max}},
\end{equation}
where $cc_\mathrm{max}$ is the maximum of coincidence rate observed in the center of the anti-dip and $cc_\mathrm{min}$ is the minimum or baseline observed for non-interfering photons due to temporal delay.

\begin{figure}
\includegraphics[scale=0.55]{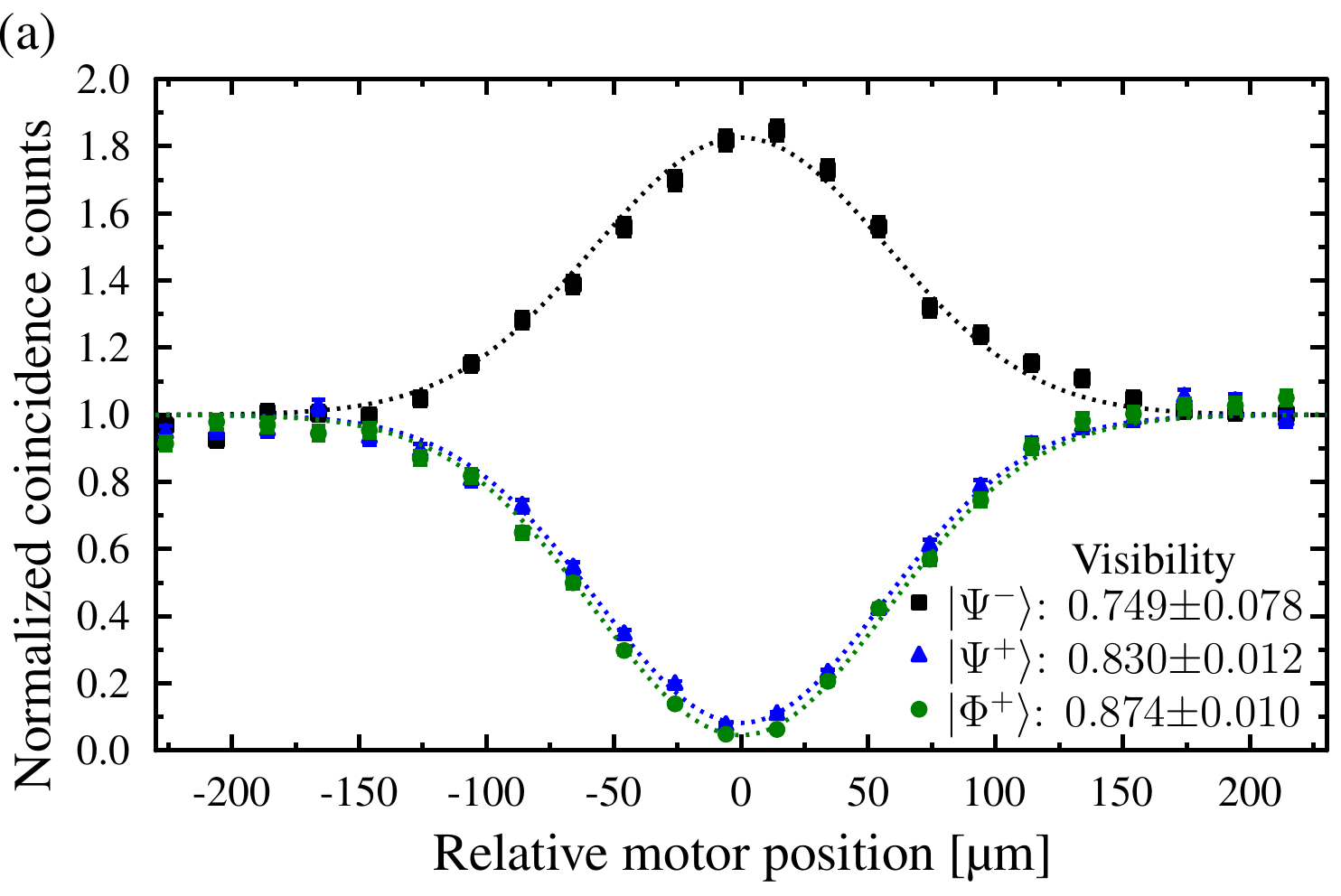}
\includegraphics[scale=0.55]{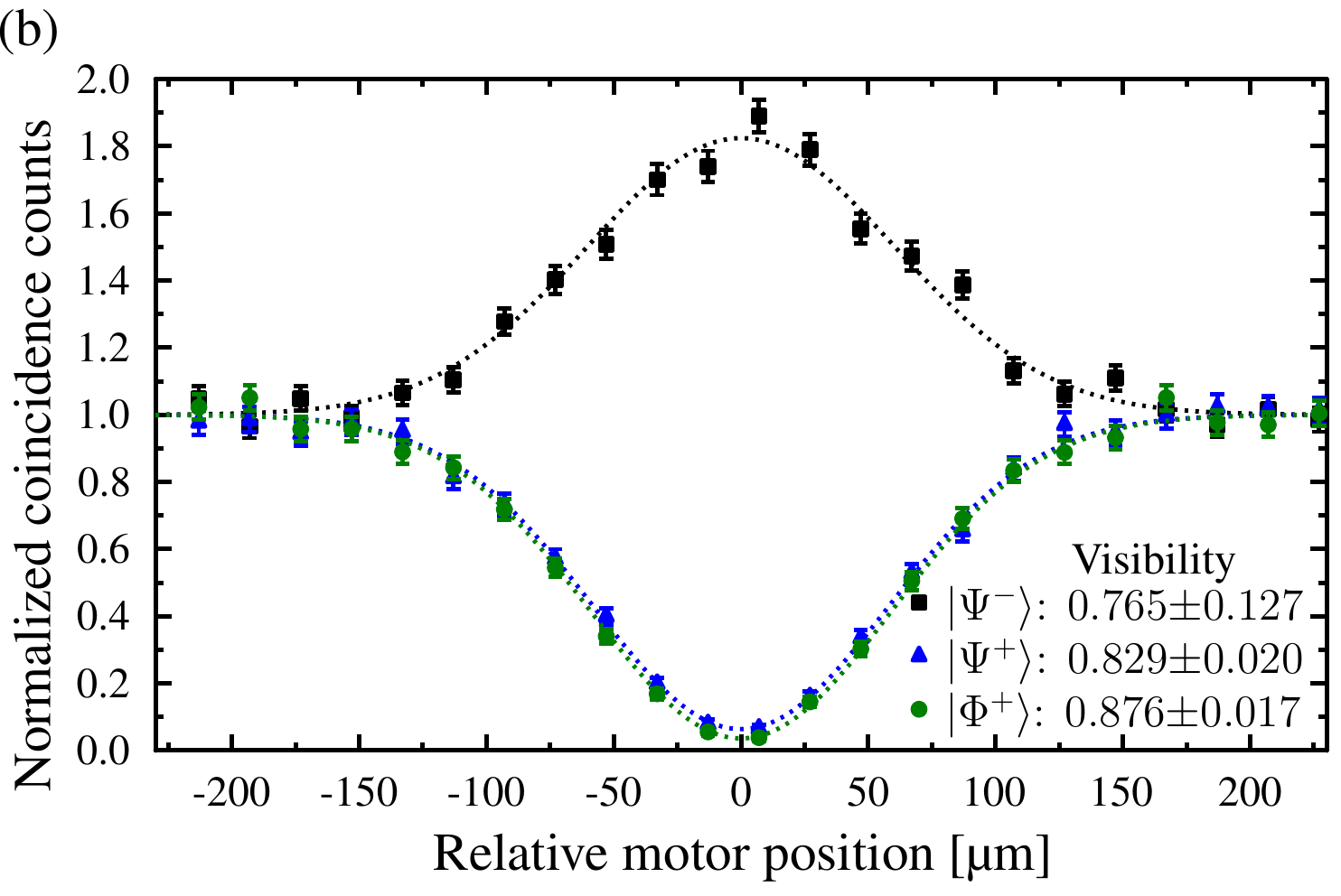}
\caption{\label{fig:HOMent} Hong-Ou-Mandel dips and anti-dips and their corresponding visibilities observed for three entangled states $|\Phi^+\rangle$ and $|\Psi^\pm\rangle$ (defined in the text) in case of (a) forward and (b) backwards propagating photon pairs.}
\end{figure}

For the purpose of thorough testing, we have set the source to generate the  same Bell states (i) $|\Phi^+\rangle$ [see definition below Eq. (\ref{eq:mixed})],  (ii) $|\Psi^+\rangle = \frac{1}{\sqrt{2}}\left(|HV\rangle + |VH\rangle\right)$ and (iii) $|\Psi^-\rangle$  in both forward and backward propagating direction. Photons in $|\Phi^+\rangle$ state bunch like two indistinguishable photons and hence this state was used only for verification of their spectral indistinguishability. On the other hand, in case of approximately $|\Psi^\pm\rangle$ states, the interference visibility depends also on the balance between the $|HV\rangle$ and $|VH\rangle$ terms and on their mutual phase. Ideally, photons in $|\Psi^+\rangle$ state completely bunch (dip) while photons in $|\Psi^-\rangle$ state anti-bunch (anti-dip). Any imperfection in the the balance between the $|HV\rangle$ and $|VH\rangle$ components or imperfection in setting of their mutual phase shift causes visibility to drop.  Results of the above mentioned interference measurement are depicted in Fig.~\ref{fig:HOMent}. Visibilities in both forward and backward directions were close to 80~\% even for $|\Psi^-\rangle$ states. During this procedure, IF (3.0) interference filters were used.

Note that for a quick check of the source operation, the $|\Psi^-\rangle$ measurement suffices. On the other hand, one can not directly identify possible causes of decreased visibility from such measurement.

\subsection{Two-photon interference between photons of different pairs}
\begin{figure}
\includegraphics[scale=0.55]{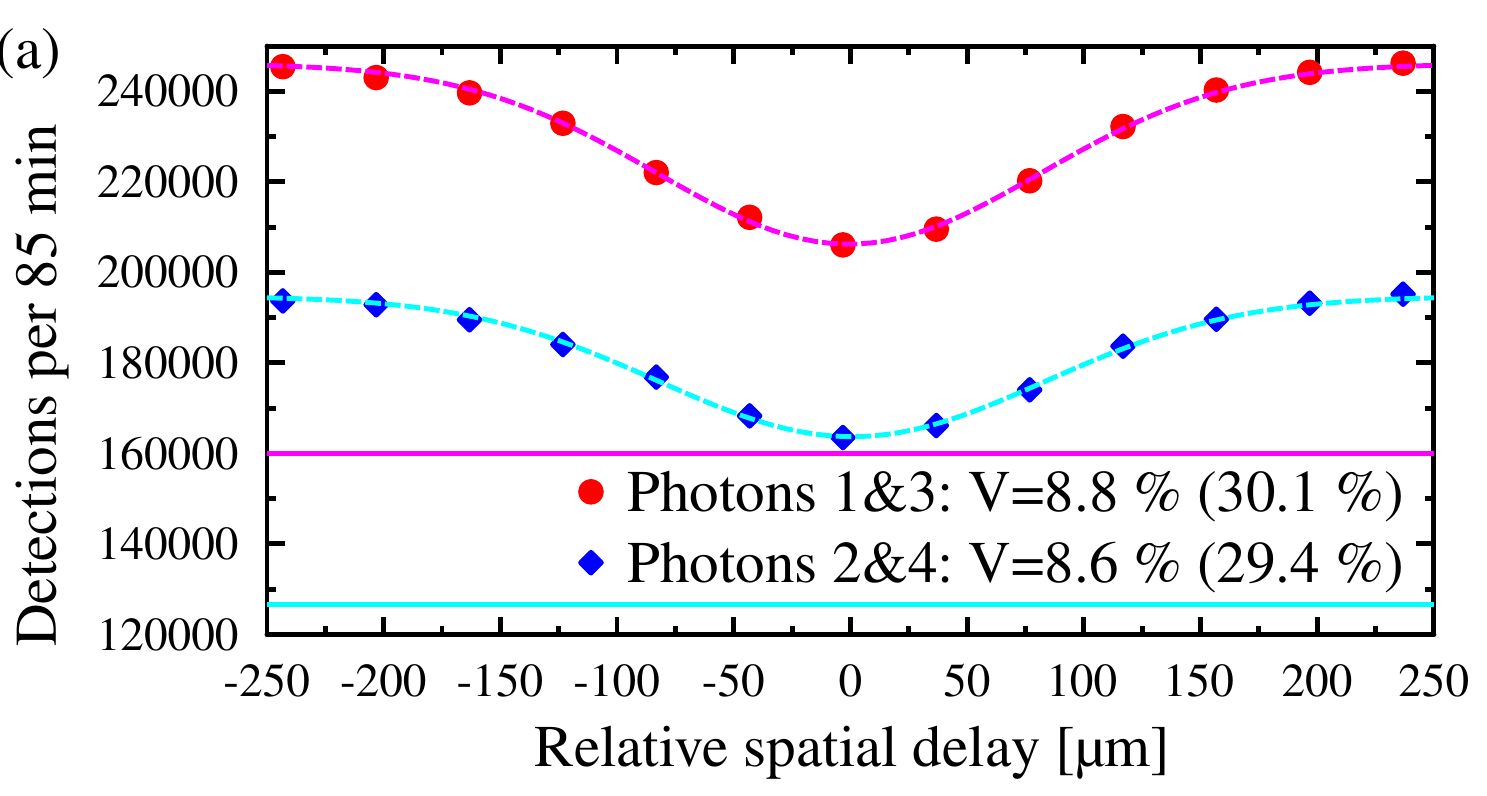}
\includegraphics[scale=0.55]{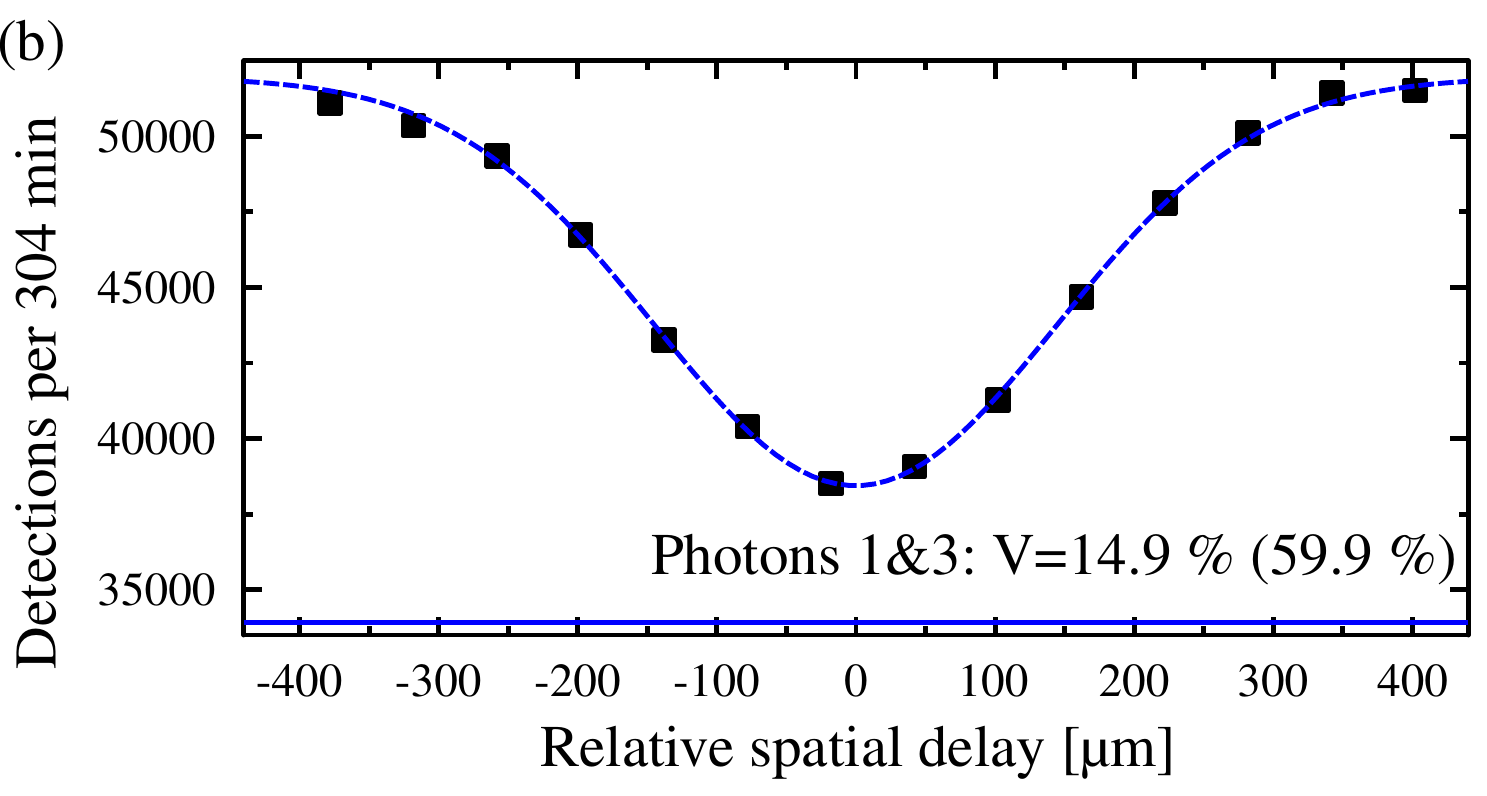}
\caption{\label{fig:HOMcross} Observed Hong-Ou-Mandel interference between photons originating in different pairs (a) IF (5.5), (b) IF (1.5). Horizontal solid lines show level of multiphoton contributions as discussed in the text. Visibilities in parentheses are calculated after subtracting these contributions.}
\end{figure}
So far we have only focused on testing the qualities of the generated photon pairs. However, complex quantum gates require interaction between more than two photons and thus require the interference also between photons originating in two different pairs.

In this configuration, we overlap one photon from the forward-generated pair with one photon from the backward-generated pair using again a 2$\times$2 fiber coupler leading to detectors (see Fig.~\ref{fig:setupHOMs}b). This time we register two-fold coincident detections of photons 1\&3 (upper modes in Fig.~\ref{fig:setup}) and also of photons 2\&4 (lower modes). Relative delay between these photons is achieved by translation of the mirror M$_3$. In contrast to the photon pairs, interference of two independent photons poses many more experimental challenges. Firstly, the photons are generated independently at any time the pumping pulse travels through the crystal. Hence, there is inherently a generation-time jitter causing imperfect overlap of the photons. Secondly, the probability of coupling one photon from the forward-generated pair and one photon from the backward-generated pair is close to the probability of coupling two photons from the forward generation direction (second-order SPDC process) and none from the backward direction or vice versa. These multi-photon contributions limit significantly the attainable visibility.

We have measured Hong-Ou-Mandel dips between the photon 1\&3 and 2\&4 using IF (5.5) filters. The raw visibilities read 8.8~\% and 8.6~\%, respectively (see Fig.~\ref{fig:HOMcross}). However, when we subtract the coincidence counts corresponding to the multi-photon contributions, values of visibilities increase to 30.1~\% and 29.4~\% which is still far from perfect. This is because of the arrival time jitter caused by the independence in photon generation. Note that the observed dips are considerably wider then expected with IF (5.5) filters. To minimize the effect of the jitter, we have repeated the same procedure with IF (1.5). With these filters, the dip is about 2.5$\times$ wider making the jitter less visible. IF (1.5) filters allowed us to reach raw visibility of 14.9~\% and after compensation on multiphoton contributions even 59.9~\%. We have accumulated the signal long enough to reach typical uncertainty of the visibility of 0.1~\%.

\subsection{Heralded two-photon interference between photons of different pairs}
\begin{figure}
\includegraphics[scale=0.55]{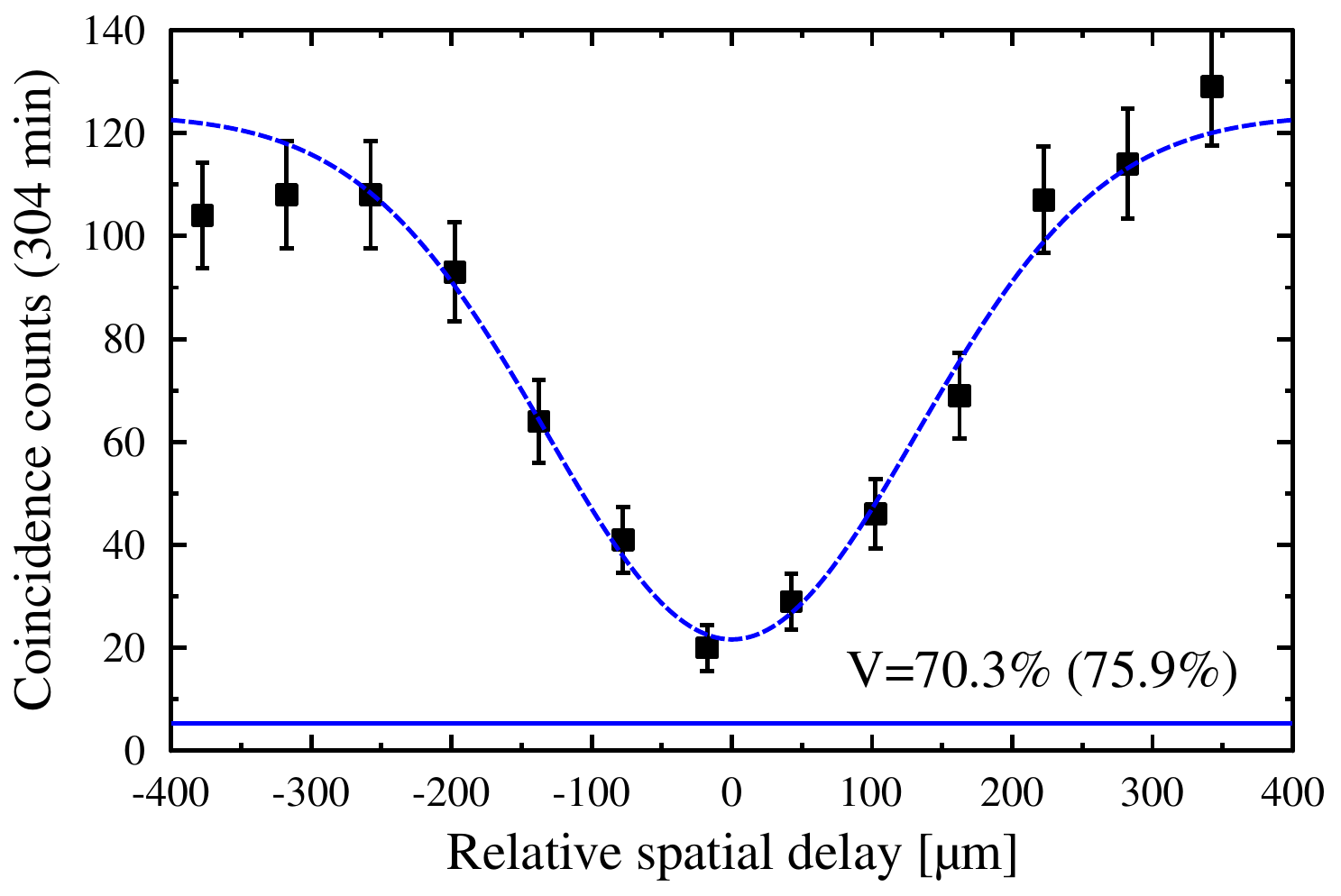}
\caption{\label{fig:HOMtrig} Observed Hong-Ou-Mandel interference between photons of different pairs (modes 1\&3) heralded by simultaneous detection of photons in modes 2 and 4. IF (1.5) was used in the overlapping modes and IF (10) in the heralding modes.}
\end{figure}

The multi-photon contributions are an inherent source of noise resulting from the photopulse statistics of the SPDC process. Generating photons in pairs  allows to circumvent this issue by proper heralding. In the configuration presented in this subsection, we have overlapped modes 1 and 3 on a fiber coupler as usual, but simultaneously, we have directed modes 2 and 4 straight to detectors (see Fig.~\ref{fig:setupHOMs}c). Detection of a photon in mode 2 heralds presence of a photon in mode 1, and similarly detection of a photon in mode 4 heralds presence of a photon in mode 3. Therefore, when registering four-fold coincidence detections in all four modes, there is a significantly reduced probability of a multi-photon contribution from any of the overlapping modes. In theory, interference visibility can reach the value of 1. The overlapping modes were equipped with IF (1.5), but the remaining generation-time jitter still reduces the observed visibility which was 70.3~\% without any corrections and 75.9~\% when the remaining multi-photon contributions were subtracted. The resulting Hong-Ou-Mandel dip is depicted in Fig.~\ref{fig:HOMtrig}.

\section{Conclusions}
In this paper, we have described in detail construction of a versatile source of four photons. To facilitate possible reproduction of our scheme, we have also provided specific adjustment and calibration procedures. To verify the correct operation of our device, we have subjected it to a thorough testing procedure. By means of a quantum state tomography, we have established that states of purities above 90~\% can be encoded on both the forward and backward propagating pairs. The source is able of generating various states including pure separable, pure entangled as well as partially and maximally mixed states. We have also included stability testing of the generated state across several hours. Indistinguishability of generated photons was tested using two-photon interference (Hong-Ou-Mandel bunching). First, we have measured interference visibility on photons belonging to the same pair. With suitable filtering, IF (3.0) or narrower, the visibility surpassed 90~\%. In the next step, we have observed visibility between photons belonging to different pairs. Due to the inherent generation-time jitter, observed visibility was quite limited and reached only 60~\% when IF (1.5) were used and multi-photon contributions were subtracted. Finally, we have set the setup for a heralded Hong-Ou-Mandel interference overlapping photons of different pairs but simultaneously heralding their presence by detection on the remaining two photons. This way, we were able to increase the interference visibility up to 76~\%. Such visibility already allows demonstrating the working principle of a number of quantum protocols. Typical rate of four-fold coincidences in this particular configuration was about 1 per minute. The presented source already proved itself in practice in several QIP experiments involving entanglement detection \cite{PhysRevA.94.052334,PhysRevA.95.030102} and controlled teleportation.

\section*{Acknowledgement}
KB and KL acknowledge financial support by the Czech Science Foundation under the project No. 16-10042Y and by Polish National Science Centre under grant DEC-2013/11/D/ST2/02638.
The authors also acknowledge the projects Nos. LO1305 and CZ.02.1.01/0.0/0.0/16\_019/0000754 of the Ministry of Education, Youth and Sports of the Czech Republic financing the infrastructure of their workplace. Data management services were kindly provides by CESNET.

%

\end{document}